# Coordinated Optimization at a Hydro Generating Plant by Software Agents

Chris Foreman, Member IEEE and Dr. Rammohan K. Ragade Ph.D., Senior Member IEEE

**Abstract** – The architecture is developed for incorporating a software agent into the existing control system of a river-based hydro generating plant. The software agent will utilize a rule-based system for ease of integration. Optimization for increased power production is presented, initially for a single generating unit. An enhancement to facilitate the constraints of river level and flow by an external user as well as the goals of corporate dispatching is then developed through the existing SCADA system. Finally, expansion to coordinate multiple hydro units at a single location is presented.

## I. INTRODUCTION

Due to the low negative environmental impact and relatively free fuel in the form of water flow available to hydro generating units, hydro power remains one of the most practical forms of green power production. Incremental increases in hydro power production directly offset carbon dioxide, nitrogen oxides, and other emissions typical of fossil-based generation in addition to the monetary returns of increased power production. Yet, this power resource is often operated with a suboptimal strategy and much research is directed towards the larger production of fossil-based plants.

What is proposed is architecture to optimize local hydro unit control with enhancements intended to coordinate unit operation with external users and multiple goals. Finally, coordination with other hydro units is developed. These enhancements achieve an enterprise level solution not currently realized by classical approaches.

*A. Previous work*
There has been much work in the area of optimization of hydro generation units. Several approaches have utilized a combination of artificial neural networks and fuzzy logic either replacing or enhancing PID control. Zhang and Yuan [1] achieve good performance by replacing conventional control with a fuzzy neural network controller on a single unit. Djukanovic et al [2] utilize a neuro-fuzzy controller with self-learning capabilities to handle generator transients. Precup et al [3] developed a Takagi-Sugeno based fuzzy controller dedicated to turbine speed control. Zhang and Zhang [4] combine adaptive fuzzy theory to existing PID control for static and dynamic improvements. Ramond et al [5] examines direct adaptive predictive control and its application to improve the performance of existing PI controls for a hydro plant. In contrast, Huang [6] explores using an ant colony system implemented by multiple software agents to determine optimal dispatching of hydro generating units. A multifaceted approach addressing the deficiencies of, and solutions for, efficient hydro generation has been presented by Wittinger [7].

While these approaches have produced good results, the use of fuzzy control and predictive models are not as modular or simple to code as a software agent structure. They also do not scale well when applying to multiple units and multiple plants [6]. Neural networks still require complex training and produce a black-box approach as opposed to software agents. Finally, most of these focus on optimizing one or a few hydro units. They do not include the architecture for handling multiple goals and multiple users as will be described in this paper. The proposed architecture will build an increasingly more advanced coordinated optimization in layers so that both the value and method of each layer can be examined.

*B. Plant Description*
The plant is the Markland Hydro Generation Facility owned by Duke Energy operating on the Ohio River near Markland, Indiana USA. The plant consists of three axial-flow Kaplan turbine generating units of approximately 25MW in size and similar configuration. The turbines are controlled by a Woodward Governor 505H control system. The plant utilizes the General Electric Fanuc iFix© supervisory control and data acquisition (SCADA) system as the control system human machine interface (HMI) and data archive. The plant coexists with a dam operated by the United States Army Corps of Engineers to accommodate river traffic and maintain a set river level. To achieve river level control, the allowed flow through the plant is currently dictated by the Corps. The head and flow available for power generation, and that of the process data in this paper, is summarized in the flow duration curve as Fig. 1.

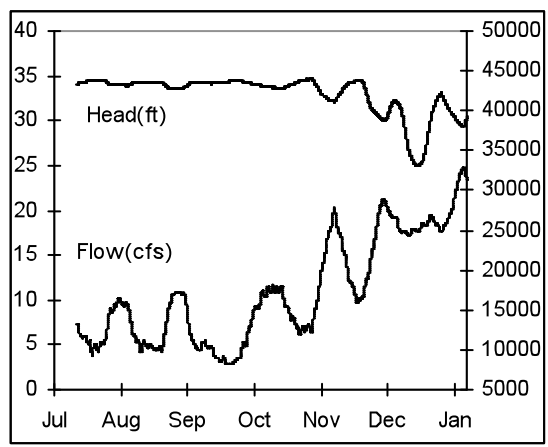

Fig. 1. Flow duration curve.

The flow of the Ohio River essentially has two annual seasons. Season one corresponds to late Spring, Summer, and early Fall and is characterized by high head and low flow. Season two corresponds to late Fall, Winter, and early Spring and is characterized by lower head and high flows. During season two, the Corps of Engineers often reduces the river level target to avoid flooding conditions upstream.

*C. Current Operating Strategy*
The Markland Hydro Generating Plant is a run-of-the-river type plant. Therefore, the plant utilizes the current flow and head to generate as much power as possible without storage or ponding considerations. The Army Corps of Engineers contacts plant operating staff and assigns an allowed flow which then becomes the plant's total flow setpoint. Operations personnel are then free to utilize this flow among the three hydro units for maximum power production and hopefully maximum profit.

## II. SINGLE AGENT ARCHITECTURE

*A. Primary Variables*
A difference in river level elevation causes water to flow from the unit entrance upstream through the turbine and out the exit downstream. This difference is the unit's net head $H_{net}$ given by (1). Together with the actual unit flow rate $Q_{act}$, the power available for generation $P_{avail}$ is given by (2)…

$$H_{net} = H_{up} - H_{down} \quad (1)$$

$$P_{avail} = k * \eta * Q_{act} * H_{net} \quad (2)$$

where $\eta$ is the unit efficiency and k is a conversion factor defined as 1 / 11810 for units of feet for head, cubic feet per second for flow, and megawatts for power. A single hydro unit is illustrated in Fig. 2. [8]

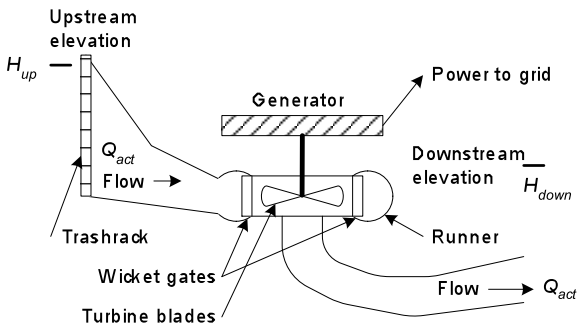

Fig. 2. Primary variables for each unit.

The variables for control actuation are wicket gate position and turbine runner blade position. The wicket gate position refers to the aperture size for river water entry into the turbine and is the main control variable for the unit's flow rate. The turbine blade position refers to the pitch of the blades from horizontal. The blade position is used to extend the efficiency of the turbine at higher gate positions since power is developed by the reaction of water pressure against the turbine runner blades. [8]

*B. Existing Control Scheme*
Control of the wicket gate position and therefore unit flow rate is accomplished by a PID loop. A PID loop includes: a linear gain component Proportional to the error; an Integral component that accumulates as the error persists in time; and a Derivative component that accounts for the rate of change in the error. A simple PID definition is given in (3). The constants $k_p$, $T_i$, and $T_d$ refer to tuning coefficients for the proportional, integral, and differential portions of the algorithm respectively. The wicket gate position $GP$ is the analog output of the PID. The error $\varepsilon$ is the difference between the user-entered flow setpoint $Q_{SP}$ and the time-varying actual flow $Q_{act}$ defined by (4). The PID acts to produce an output $GP$ that reduces the error signal $\varepsilon$ to zero.

$$GP = k_p \left( \varepsilon + \frac{1}{T_i} \int (\varepsilon) dt + T_d \frac{d}{dt}(\varepsilon) \right) \quad (3)$$

$$\varepsilon = (Q_{SP} - Q_{act}) \quad (4)$$

Control of the turbine blade position $BP$ is by a software cam. A software cam is modeled by a virtual 3-dimensional surface where independent variables X and Y are mapped to a dependent variable Z. In this case, X and Y refer to gate position and net head while Z refers to the blade position determined from these inputs. The software cam is explicitly and statically defined from an index test which is performed at a set of given heads and flows and then linearly interpolated during use. The concept of a software cam originates from the mechanical linkages used between the gate position and blade position with a hardware cam before the existence of computer-based control.

The current control scheme for calculating the gate and blade positions is shown in Fig. 3 simplified by the omission of safety systems, supporting equipment, and other balance-of-plant control. These are calculated in parallel and in real-time as control loops.

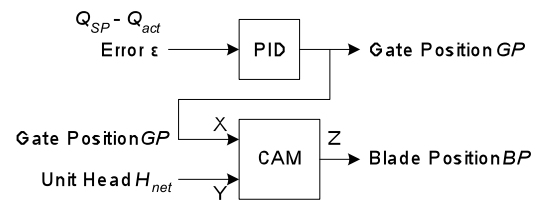

Fig. 3. Current control scheme.

*C. Incorporation of a Software Agent for a Single Unit*
A single software agent is then added to the control scheme of Fig. 3 within the existing control software. This agent will determine the appropriate actions for the control variables $GP$ and $BP$. This builds a foundation for optimal control and provides enhancement for multiple user goals. These actions are incorporated as in Fig. 4.

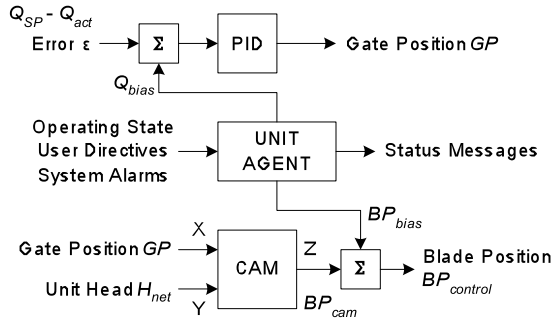

Fig 4. Individual unit integration.

The two primary degrees of influence for unit operation are unit flow biasing and blade position biasing. The unit flow biasing is accomplished by adding the software agent's calculated value $Q_{bias}$ to the error signal to affect the user-entered flow setpoint $Q_{SP}$. The blade position biasing is accomplished by adding the software agent's calculated value $BP_{bias}$ to the output of the software cam $BP_{cam}$ to produce a shifted value used for control $BP_{control}$. The final actions are summarized in (5) and (6).

$$\varepsilon + Q_{bias} = (Q_{SP} + Q_{bias} - Q_{act}) \tag{5}$$

$$BP_{control} = (BP_{cam} + BP_{bias}) \tag{6}$$

The *operating state* refers to the set of control system variables that define the current operating state of the unit. This includes variables already mentioned here in addition to river water temperatures, current power output, and other variables that may be relevant to determine operating state. *User directives* refer to conditions set by the users of the control system. Users include unit operators, corporate managers, and other software agents as will be shown later that may place limits on, or demand new goals for, unit operation. *System alarms* refer to over-temperature conditions, vibration / cavitation problems, and other trouble conditions that may require intervention by the unit agent to correct. The output of the unit agent is shown as status messages. These include the agent's status for reporting to unit operators and managers as well as messages to other unit agents when multiple units are to be coordinated, discussed later.

D. *Architecture of a Single Unit Agent*
The base optimization engine used in the agent is a rule-based system. The use of such a system provides several benefits.
1. The agent is a white box solution. A rule-based system allows engineers to see inside the agent and understand how it is functioning. This simplifies troubleshooting which is a significant issue in process control.
2. The agent is a portable solution. A rule-based system can be constructed and simulated with readily available environments such as: XML; CLIPS and variants; the Java Expert System Shell JESS; etc. These environments often allow the developed system to be exported as a set of IF-THEN and similar statements that are easily integrated into almost any vintage control system.
3. The agent is a modular solution. A rule-based system can be easily expanded to handle additional tasks by adding more rules. The software agent architecture further facilitates the modular approach via multiple agent instances.
4. The agent is a multi-goal solution. A rule-based system naturally balances multiple users and multiple goals through rule negotiation where artificial neural networks and even classical approaches are often single minded.

Fig. 5 illustrates the architecture of a single agent.

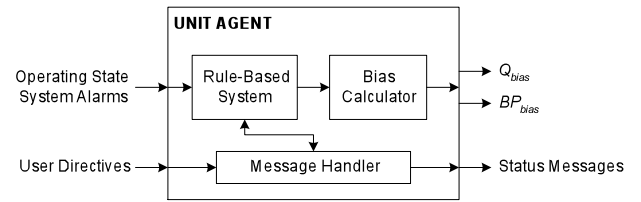

Fig. 5. Agent architecture.

The rule-based system block takes the process variables in the operating state and the system alarm conditions into account to produce a response for the bias calculator block. The bias calculator block computes real-time numeric values for the biases to be incorporated into the existing control scheme per Fig. 4. The message handler block accepts messages from other users and agents to influence the rule-based system block as well. The message handler also generates messages to respond to these users and agents. Fig. 6 illustrates an activity diagram of a typical bias calculation and thus optimization step.

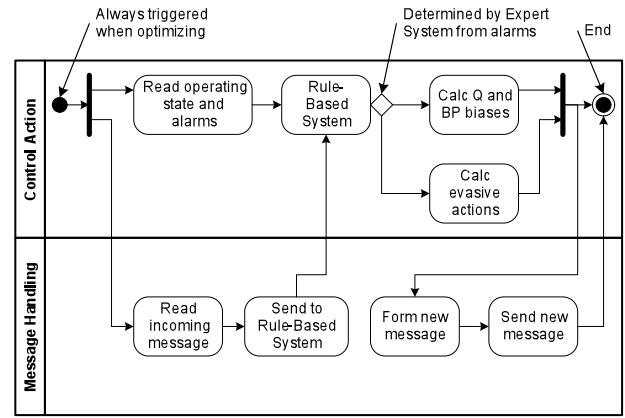

Fig. 6. Activity diagram of single agent optimization.

The activity diagram of Fig. 6 illustrates a single optimization step divided into swim lanes to illustrate the control action and message handling segments of the agent. This activity is executed in the real-time environment of the control system and is realized as a continuous loop during execution. Therefore, the loop is retriggered at the end point either immediately or after a delay for settling time of the process. Depending on current conditions, the agent can either continue calculating optimization biases or take other predefined evasive actions to address the alarm states as dictated by additional rules in the rule-based system block. The details of the message handler and the format of messages will follow in later sections. Here they are generically referred to in order to illustrate their role in the optimization process. In the following sections, the rule-based system block will be enhanced with additional rules to utilize messages from other users and agents and report its own actions to these entities.

## III. ENTERPRISE ARCHITECTURE

The single agent architecture of section II is now enhanced to allow multiple users to interact with the agent. This is necessary since the operation of the hydro generating unit also relies on the strategic input of various users that are assigned influence over the unit. The two users considered are the Army Corps of Engineers and the corporate dispatching office.

### A. Strategies and Goals of the Users

The Army Corps of Engineers is tasked with maintaining the upstream river elevation within a one-foot tolerance of 455ft above sea level and locking river traffic through the dam. To maintain upstream elevation, the corps uses manual look-up charts based on upstream conditions to determine the flow necessary to maintain the upstream level. The corps also reserves 5000CFS of available water flow for locking purposes all the time. Once the corps determines the river flow requirements and subtracts the locking requirements, the resulting value for available flow is manually reported to the hydro plant via a SCADA interface. This value becomes $Q_{SP}$ mentioned above in (4) since the plant must utilize all of this flow and no more or risk interfering with the Corps' control of the upstream river elevation.

The corporate dispatching office is tasked with dispatching generating units in the corporate fleet to meet customer demand and maintain stability of the power delivery grid. Because the hydro units utilize the cheapest source of fuel, being river water flow, and have the least environmental impact in terms of pollution, they are normally best dispatched to deliver the most power generation all the time. When demand is low, keeping the hydro units at maximum generation allows fossil-fueled units to be backed off reducing fuel costs and emissions. The corporate dispatching office does occasionally need to adjust power delivery for grid stability issues and would also benefit from the unit status updates the software agent could provide.

### B. Communicating with the Unit Agent

Process control systems already include graphical user interfaces referred to as HMIs. In this case, a SCADA system serves as the HMI as is typical of remote control. The plant operator, corporate dispatcher, and Army Corps of Engineers each have a SCADA interface. While the plant operator is responsible for operating the unit, corporate dispatch monitors generation and the Corps sends the available flow to become the setpoint $Q_{SP}$ as mentioned earlier. The SCADA interface provides a different level of security for each user to control that user's access to protected data and restrict certain control actions. Interaction with the unit agent is through its message handler via messages. The format of messages is given in Fig. 7.

Message

| ID: \<number\> | User: \<name\> | Status: \<list\> |
|---|---|---|
| e.g. | e.g. | e.g. |
| • 1 | • operator | • $Q_{bias}$= +1000CFS |
|  | • corps | • $BP_{bias}$= 10% |
|  | • dispatch | • stator over temp |
|  | • unit1 | • new $Q_{SP}$ |
|  |  | • load shed |
|  |  | • disable agent |

Fig. 7. Message format.

A message *ID* is used for tracking purposes and to maintain chronological order for processing. The *User* field indicates the originating user or agent. If the user is a unit agent, the *Status* field will list the biases and any local alarms such as stator temperature high. Otherwise, commands from other users such as: the Corps issuing a new flow setpoint; dispatch commanding the unit to reduce load; or the operator disabling the agent's influence will be listed.

To facilitate messaging, each agent can have a control system process point defined for their use. Process points are information structures predefined by all process control systems consisting of multiple data fields. Communication is accomplished by other users reading from certain data fields to get status from the agent and writing to other fields to send information to the agent. Using a predefined structure built into the existing control system provides a simple and established communications interface. Fig. 8 illustrates a use case diagram for the operator, Corps, and dispatch users as they interact with the unit agent.

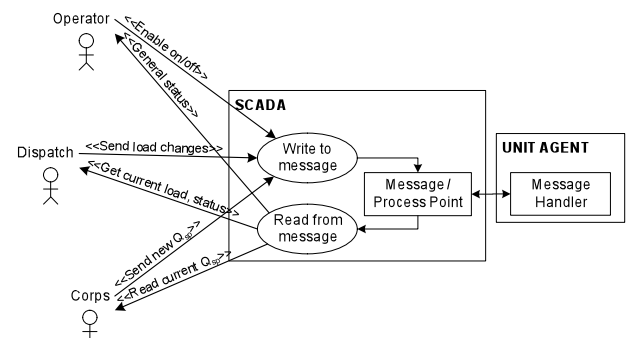

Fig. 8. Use case diagram.

## IV. MULTI-AGENT ARCHITECTURE

Since this hydro plant houses three generating units, the optimal operating solution will require the coordination of these units. Therefore, the architecture is further enhanced to allow the individual unit agents to be coordinated. Since the existing control system did not provide for coordinating multiple units, enhancing the agents with this ability will achieve this coordination goal without additional configuration of the control system.

*A. Strategies and Goals of Multiple Agents*

The single unit agent defined previously now acquires additional users in the form of other instances of this agent type. This agent also becomes a user of these other instances as well, resulting in a peer network for multiple units. An expanded use case diagram is shown in Fig. 9 illustrating this.

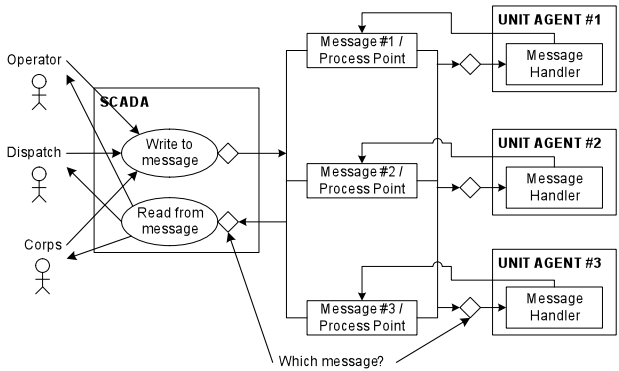

Fig. 9. Multi-Agent use case diagram.

In Fig. 9, the roles of the users utilizing the SCADA system remain as in Fig. 8 but are omitted here for clarity. Decision blocks are added to indicate a choice of which message process point to read or write. For the SCADA users, the user makes this choice with the SCADA system by manual selection. In the case of unit agents, they always write to their associated message process point but can read from any of the others. The decision of which message process points to read is determined by the definition of the rules in that agent's rule-based system block. Therefore, rules that need information about another agent automatically fetch the desired information.

Each unit agent attempts the optimal production of power for its unit while coordinating the directives of other users. With the addition of other unit agents as users, the status of other generating units is now able to influence this unit agent. This influence is handled by again expanding the set of rules in the rule-based system block. For example, unit agent #1 reports a high temperature alarm and must thus reduce its unit's generation to reduce the demand on the stator until it can cool sufficiently. This requires unit #1 to give up some water flow. Unit agent #2 receives this status and increases its unit's flow setpoint to take advantage of the available water and generate more power.

## V. IMPLEMENTATION

*A. Motivation and Expected Results*

The original control was remotely by operators via the SCADA system. While the software cams defining blade position had been previously programmed for each unit, they were still statically defined and did not account for changes in unit conditions or special circumstances. If the operators had free time, they would sometimes manually adjust the blade position and discover additional generating capacity with the existing head and flow rate. This indicated that hidden generating capacity was available if one were to intelligently search considering the current unit conditions. The use of software agents were selected as the best candidate for this task. A sample case is illustrated in Fig. 10 for a single unit.

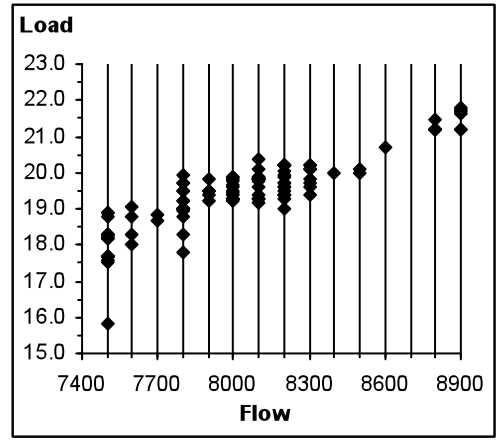

Fig. 10. Load vs Flow.

In this case, the time range selected was from July through September, being the range of relatively consistent head depicted in Fig. 1. Unit net head $H_{net}$ was selected at approximately 34ft as in (1). Each point of the data is a cluster of similar steady-state patterns with a minimal support of 15 adjacent patterns or 15 minutes. This figure depicts multiple power production (load) points for a given flow. With all other values being similar, the major difference is in blade position. Assuming the highest point for a given flow value represents an estimate of potential power generation while the median value represents a typical blade choice, a conservative estimate of at least 0.5MW of increased instantaneous power generation is achieved. Assuming this increase of 0.5MW for 50% of the time for one year, results in 2190MWhr of additional power generation. This offsets 912.5tons of less coal on an annual basis by (7). This also reduces annual carbon dioxide release approximately 1670tons by (8). This is compared with a typical coal fired generating unit operating with a heat rate (HR) of 10MBTU/MWhr. Fuel is assumed as bituminous coal with a higher heating value (HHV) of 24MBTU/ton and 75% carbon composition.

$$\text{Coal Tons/yr} = \text{Power} * HR / HHV \quad (7)$$

$$CO_2 \text{ Tons/yr} = \text{Coal Tons/yr} * 1.83 C/CO_2 * 75\% \quad (8)$$

This justifies the effort for optimization and the potential for greater generation plus a more user friendly process is achieved through the use of an agent-based system.

*B. Process Data Collection*

Process data for 200 days was extracted from the SCADA system in the form of comma-separated-variable text files. Each line or pattern from the file represents a time-stamped one minute sample of the collected variables for all units. A pattern consists of the following variables for each of the three units as previously defined:
1. Gate Position $GP$
2. Blade Position $BP$
3. Net Head $H_{net}$
4. Actual Unit Flow $Q_{act}$
5. Allocated Unit Flow $Q_{SP}$
6. Power Production (Load) $P$
7. Stator Temperatures
8. Vibration Measurements

The pattern also contains the following plant-wide variables:
1. Plant Head $H_{net}$
2. Plant Actual Flow = $\Sigma Q_{act}$
3. Plant Allocated Flow = $\Sigma Q_{SP}$
4. Plant Power Production (Load) = $\Sigma P$

Alarm conditions are extracted from a log file that is time correlated with this process data.

*C. Integration into the Plant Control System*

Integration into the plant control system is achieved by building an initial rule-based engine for each unit, defining the process points for message handling, and building the graphical user interface (GUI) in the HMI for the SCADA users to interact with. These software artifacts are intended to be implemented within the existing control system hardware.

The rule-based engine is built externally and then exported as IF-THEN statements, SELECT-CASE statements, conditional loops, or similar. These statements are then coded directly into the control system. Some rules are already known by engineering staff from current experience. Mining of the collected data for hidden relations has lead to additional rule ideas, particularly from the collected data that captures actions taken by the operators for specific detectable events.

Process points for message handling are pre-existing data structures in the control system. The control system infrastructure automatically handles the network communications and timing for these message points as well. Therefore, their definition becomes trivial with a multi-attribute point reserved for each agent. For example, an agent reporting its $Q_{bias}$ would write this numerical value to its message point's corresponding attribute data field in a *point.attribute* format as in (9).

$$Q_{bias} \text{ value} \rightarrow Unit1Agent.Qbias \quad (9)$$

Another user or agent interested in this information as a rule condition would access it by reading this point attribute in a similar manner.

The HMI for users to interact with the agents is implemented with the existing SCADA system. The interface will vary depending on which user is accessing the system. This is determined by user id. From the use case diagram in Fig. 8, the three users utilizing the SCADA HMI are the unit operator, the corporate dispatching office, and the Army Corps of Engineers. Samples of the HMI screens are given in Fig. 11. The minimum attributes implemented on the HMI screens by user are as follows:
1. For the Unit Operator…
   a. The ability to manually enable or disable the unit agents.
   b. The ability to observe the direct actions the unit agents are taking to influence the process.
   c. The ability to observe the estimated benefit of the actions.
   d. The ability to observe any relevant alarms and trouble conditions.
2. For Corporate Dispatching…
   a. The ability to enter a new load setpoint.
   b. The ability to observe the current power load.
   c. The ability to observe the estimated benefit of the actions.
   d. The ability to observe any relevant alarms and trouble conditions.
3. Corps – The Army Corps of Engineers
   a. The ability to set the flow setpoint $Q_{SP}$ for the whole plant.
   b. The ability to observe the current flow $Q_{act}$.

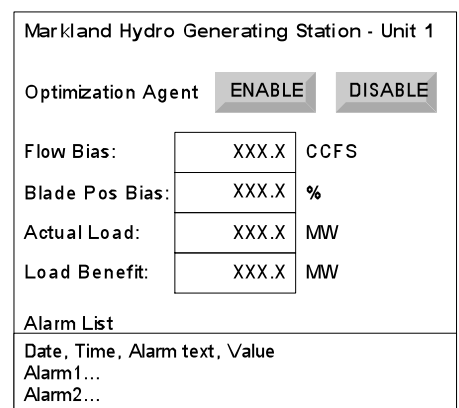

Fig. 11a. Operator's HMI SCADA screen.

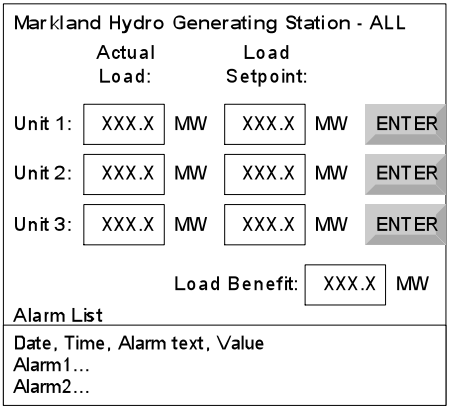

Fig. 11b. Dispatch's HMI SCADA screen.

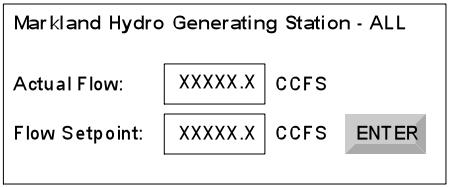

Fig. 11c. Corps' HMI SCADA screen.

### D. Optimal States of Operation

In the architecture, the state of an individual generating unit's agent is determined by multiple users and conditions. Corporate Dispatching will be submitting generation goals, the Corps of Engineers will be submitting plant flow goals, and the unit agents themselves will be modifying gate and blade positions to obtain the maximum load benefit when not handling other users' directives. Alternatively, when unit trouble conditions exist, the unit agent will be modifying gate and blade positions to correct the trouble condition and these corrections may take precedence over the directives of the other users. This allows the model to continuously search for the best solution within its rule-set capability while handling multiple goals and constraints. A simple state space diagram is illustrated in Fig. 12.

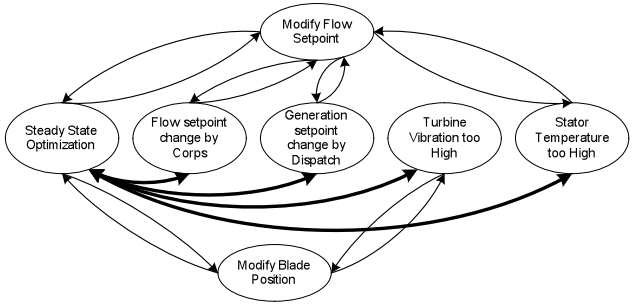

Fig. 12. Agent State Space.

From Fig. 12, there are five major states depicted depending on the current conditions and user directives. There are two minor states of modifying the flow setpoint and blade position once new biases have been calculated by the agent, per (5) and (6). If there are no current directives or trouble conditions, the agent resides in the first state of steady state optimization biasing flow setpoint and blade position. Transition from this steady state, depicted by bold lines, to one of the other main states is therefore when a new directive or trouble condition occurs. From any of the main states, the agent can modify flow setpoint or blade position, as applicable, to achieve the goal of that state. For example, changing plant flow setpoint, generation setpoint, and handling temperature problems are accomplished primarily by biasing flow setpoint position while handling turbine vibration / cavitation is accomplished by modifying blade position.

## VI. RESULTS

The data referenced in the implementation section is used to build and verify a mathematical model of the hydro units. The use of such a model allows development to progress without interruption to the plant. The agents are developed with this model and the optimization is evaluated on the following criteria:
1. Achieving additional load within an existing flow setpoint, thus improving production.
2. Achieving additional coordination of Operator, Dispatch, and Corps users, thus improving strategic operation.
3. Handling trouble conditions in an automated manner, thus improving stability and reliability.

Rules used in the agents would be coded by the plant's control system engineer. Required rules for stable operation as well as some desired operating guidelines would come from plant staff. New rules may always be added dynamically to the architecture from the results of data mining or operators' perspectives in various situations.

### A. Hydro Unit Model Definition

The hydro unit is characterized by clustering the collected data patterns, discussed in the Process Data Collection section, to build a database of operating states. The independent input variables for this clustering are net head $H_{net}$ and flow setpoint $Q_{SP}$. The dependent variables are gate position $GP$, which is a function of flow defined by (3,4), and blade position $BP$, which is a function of $H_{net}$, and $GP$ by Fig. 3. Power $P$, or load, is also defined by $H_{net}$ and $Q_{act}$ and efficiency $\eta$ as in (2). $BP$ modification is used to improve $\eta$ and thus in relative terms $\eta$ is largely a function of $BP$. Efficiency can then be determined from a known state in the database and then used to solve for an unknown state near the known state with the same $BP$. Using the above equations, states between these can be interpolated or extrapolated to determine probable results.

### B. Achieving the Goals of the Criteria

With a model in place, rules can be developed and implemented in agents for evaluation. For this evaluation and due to the difficulty in obtaining access to the generating unit, the results are simulated based on the collected data. This simulated method is actually superior because it allows risk free analysis and avoids disturbing the generating process until a probable theory of operation is determined. Below are the methods of this optimization followed by actual samples from operating data.

The feasibility of achieving additional load within an existing flow setpoint is already demonstrated in Fig. 10 of the Motivation and Expected Results section. This goal would be achieved by a rule set in the agent that looks for operating states in the database that produce a higher power output with similar head and flow values. The controllable parameter of blade position would be biased to move to this new state. Gate position *GP* would be automatically adjusted by the PID to maintain flow. For example, to increase the torque against the turbine blades for maximum production, find in the domain ***D*** of operating state clusters…

$$D\ (H_{net},\ Q_{sp}) \rightarrow BP_{control} \ni P_{opt} > P_{act} \quad (10)$$

then calculate…

$$BP_{bias} = BP_{control} - BP_{cam} \quad (11)$$

where $P_{opt}$ and $P_{act}$ are the power before and after the optimization step respectively.

Another example would be to bias both $Q$ and $BP$ position of one unit for increased generation at the expense of another less efficient unit to achieve a net gain in plant generation at the same total plant flow. From (2) we have…

$$P_{act}(H_{net},\ Q_{sp},\ \eta) \quad (12)$$

noting that efficiency $\eta$ is a function of *BP* for a given head and flow leads to…

$$P_{act}(H_{net},\ Q_{sp},\ BP_{control}) \quad (13)$$

For multiple generating units find, as before from $H_{net}$ and $Q_{sp}$ …

$$(Q_{sp_i}, BP_i) \ni \sum_i P_{opt_i} > \sum_i P_{act_i} \quad (14)$$

then calculate the appropriate $Q_{bias}$ and $BP_{bias}$ similar to (11). This is demonstrated in Fig. 13.

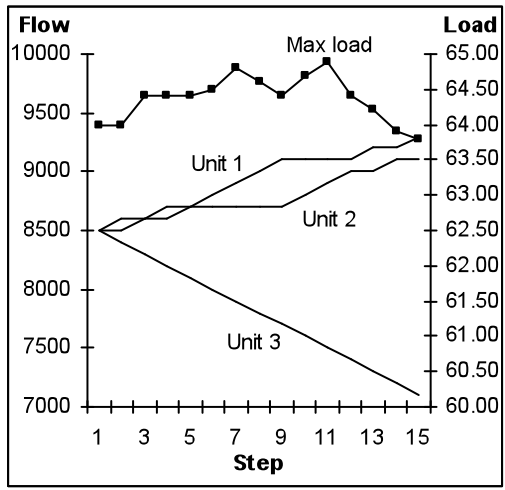

Fig. 13. Load vs Flow Redistribution.

Three generating units start at an equal flow. Based on their individual efficiencies from (2), flow is incrementally redistributed with priority given to the most efficient unit. Notice that the maximum load line peaks at a higher value before trailing off as expected from a diminishing returns function. Since there will always be a least efficient unit, flow is inevitably taken from this unit until its efficiency drops, resulting in a net loss for the plant. Therefore, the agent needs to detect this and cease flow redistribution prior to this point. The gain depicted in Fig. 13 represents nearly 1MW of additional power generation.

One of the controllable variables influencing unit efficiency is quantified by *drawdown*. Noting Fig. 2, the trash rack is a screen type filter that keeps river trash from passing through the unit and damaging the turbine blades. Drawdown is a distance measure that reduces the upstream head $H_{up}$ and thus reduces $H_{net}$ due to the restriction in flow caused by accumulated trash. This trash is released by performing a *load eject* which shuts the unit down abruptly causing a back flow to push the trash away. Since there is a penalty associated with this action, there are optimal times to do this. The loss of power associated with trash accumulation is derived from (2) and given by (15) where $Q_{act}'$ and $H_{net}'$ are the reduced $Q_{act}$ and $H_{net}$.

$$P_{loss} = k * \eta * [Q_{act} * H_{net} - Q_{act}' * H_{net}'] \quad (15)$$

A rule in the unit agent can monitor the real-time value for $P_{loss}$ and perform a load eject at some threshold level dependent on the current unit generation and balanced with the users' goals in effect. In general, this would be when the projected gain in generation from reducing the drawdown is greater than the cost $P_{loss}$ of performing the load eject.

The goal of achieving additional coordination of Operator, Dispatch, and Corps users is largely qualitative. Rather than rely on manual reporting to change generation and allowable plant flow, this information can be integrated into the control model automatically at each user's convenience. More detailed status becomes available in real time to the users whenever they request it. For example, one critical control parameter is allowable plant flow set by the Corps because this determines the plant generating potential. Accurate and timely updates of this parameter allow the plant to make use of all the flow allowed. The historical data, by its nature, did not record these missed opportunities since they previously occurred between Corps and Operator personnel. For comparison, one hour of missed flow at 1000cfs costs 2.55MWhr in potential generation from (2) assuming median values of $H_{net}$ = 33.5ft and $\eta$ = 90%.

The goal of handling trouble conditions in an automated manner relieves the operator from manually handling these conditions. This results in earlier implementation and better quantification of resolutions. For example, previously when a unit's stator overheated, $Q_{SP}$ was reduced until the resulting power $P_{act}$ was reduced by 5MW until the condition normalized. A rule set can be defined that only reduces power by the necessary amount to alleviate this condition thus preserving some of this generation such as…

```
While StatorTemp = HI Do {
    Q_bias = Q_bias – 500cfs
    Wait (1minute)  } #to see if stator is cooling
```

Fig. 14 illustrates a simulated handling of this condition as compared to typical operator response for a single unit. This does not account for the additional gains possible by having the agent reallocate that flow to other units during the trouble condition such that,

$$\sum_i Q_{bias_i} = 0 \tag{16}$$

and automatically restore that flow when the trouble condition normalizes, returning to steady state optimization per Fig. 12.

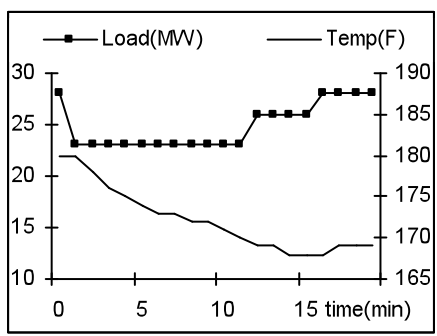

Fig. 14a. Operator response.

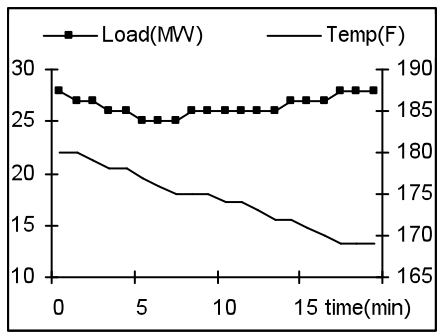

Fig. 14b. Optimized response.

In the 200 days of data, there were 71 high stator temperature events (>180degF and >10min) for unit 1, 24 such events for unit 2, and 199 such events for unit 3. Assuming 0.55MWhr gain per event as simulated in Fig. 13, this results in 161.7MWhrs of additional generation.

The similar case exists for vibration and cavitation trouble conditions utilizing $BP_{bias}$ or $Q_{bias}$. During cavitation and vibration events, the blade position can be adjusted or the flow biased to alleviate the event. A measured response can be applied rather than a step change load reduction and any flow reductions can be added to the other units as in (16).

## VII. CONCLUSIONS AND FUTURE DIRECTIONS

The results clearly show that there is free power generation left on the table due to suboptimal gate and blade positions. Furthermore, there are significant qualitative gains in raw generation as well as operating strategy by coordinating the multiple users and multiple units with intelligent agents. These agents facilitate an autonomy that, being a white-box rule-based architecture, allows simple integration and modification. This approach provides the heuristic capability necessary given the multiple layers of optimization, the multiple users utilizing the optimization, and the proposed enhancements. Some of the qualitative benefits and future directions include:

1. Through better management and response to trouble conditions, maintenance and downtime are both expected to be reduced. This translates into reduced operating costs and increased production and availability. There are also operating states that dramatically reduce cumulative machine wear under variable conditions, similar to those in this paper, studied by March [9].
2. A major investment both in money and effort is in the licensing process. Demonstrating better use of the natural resource of river flow, as well as better plant management to increase generation thus reducing reliance on fossil fuels, provides solid evidence to both licensing boards and the general public for the case of continued operation.
3. In the future, it would be mutually beneficial to the Corps and the generating company to automate control of river level for Corps. This would allow the Corps to set a target river level rather than determine the target flow to indirectly control level. This simplifies the Corps' duties and provides more accurate calculation of $Q_{sp}$, which is a critical variable for plant operation.
4. As wildlife and environmental concerns become increasingly important, the issue of fish mortality through the turbine blades must be addressed. Research has been done on models for safe fish passage by varying operating conditions that would integrate well into this model. Simple revisions to the rule sets could accomplish this. [10,11]

REFERENCES

[1] Zipeng Zhang and Xiaohui Yuan, "Research on Hydro Electric Generating Unit Controller Based on Fuzzy Neural Newtork," *Proceedings of the 6th World Congress on Intelligent Control and Automation*, pp. 6559-6563, June 2006.
[2] Miodrag B. Djukanovic, Milan S. Calovic, Bogdan V. Vesovic, and Dejan J. Sobajic, "Neuro-Fuzzy Controller of Low Head Hydropower Plants Using Adaptive-Network Based Fuzzy Inference System," *IEEE Transactions on Energy Conversion*, vol. 12, no. 4, pp. 375-381, Dec 1997.
[3] Radu-Emil Precup, Zsuzsa Preitl, and Stefan Kilyeni, "Fuzzy Control Solution for Hydro Turbine Generators," *International Conference on Control and Automation*, pp. 83-88, Jun 2005.

[4] Xiao-Ying Zhang and Ming-Guang Zhang, "An Adaptive Fuzzy PID Control of Hydro-Turbine Governor," *Proceedings of the Fifth International Conference on Machine Learning and Cybernetics*, pp. 325-329, Aug 2006.

[5] Gerard Ramond, Didier Dumur, Antoine Libaux, Patrick Boucher, "Direct Adaptive Predictive Control of an Hydro-Electric Plant," *Proceedings of the International Conference on Control Applications*, pp. 606-611, Sep 2001.

[6] Shyh-Jier Huang, "Enhancement of Hydroelectric Generation Scheduling Using Ant Colony System Based Optimization Approaches," IEEE Transactions on Energy Conversion, vol. 16, no. 3, pp. 296-301, Sep 2001.

[7] Rodney J. Wittenger, "Optimizing the Corps' Hydroelectric Generation on the Colombia River: A Multi-Faceted Effort", *Waterpower XIII Conference*, Buffalo, NY, Jul. 2003, pp. 80-89.

[8] Kermit Paul Jr. et al, *The Guide to Hydropower Mechanical Design.* Kansas City, MO, HCI Publications, pp. 3.1-3.8, 1996.

[9] Patrick A. March, "Quantifying the Maintenance Costs Associated with Variable Operating Conditions", *Waterpower XIII Conference*, Buffalo, NY, Jul. 2003, pp. 98-107.

[10] R. K. Fisher, S. Brown, and D. Mathur, "The Importance of the Operation of a Kaplan Turbine on Fish Survivability", *Proceedings of the International Conference on Hydropower*, American Society of Civil Engineers, Atlanta, GA, Aug. 1997, pp. 392-401.

[11] Steven F. Railsback and Douglas A. Dixon, "Individual-Based Fish Models: Ready for Business in Hydro Licensing", *Waterpower XIII Conference*, Buffalo, NY, Jul. 2003, pp. 90-97.

**Chris Foreman** (Member IEEE 2004) has been a controls engineer for industrial processes since 1993. He has worked primarily in the power generation industry but also has experience in manufacturing, material handling, and semiconductor processing in a cleanroom environment. Specializing in advanced control techniques and process optimization, he has managed several projects to improve production, efficiency, and reduce emissions. He has worked with companies such as Westinghouse Process Control Division (now Emerson Process Management), Cinergy (now Duke Energy), and Alcoa Inc. among others. he graduated with a Bachelor of Science in Electrical Engineering in 1990, a Master of Engineering in Electrical Engineering in 1996, and is currently pursuing a PhD in Computer Science and Engineering, all at the University of Louisville.

**Rammohan K. Ragade** (Ph.D., I. I. T. Kanpur, India (1968)) is a Professor of Computer Engineering and Computer Science at the University of Louisville. He holds a BE degree in Electrical Power Engineering from I. I. Sc. Bangalore, India (1964). He served as the Coordinator for the Ph.D. Program in Computer Science and Engineering from 1999-2005. He has written well over 100 papers, including journal articles, refereed conference papers, chapter contributions to books and is the co-editor of 4 books. He is a senior member of IEEE. He is a member of the ACM. He has taught graduate courses in Software Engineering and Advanced Software Engineering, Software Design, Computer Security, Knowledge Engineering, Computer Architecture, and Simulation Modeling. His research interests include agent technologies, object oriented methodologies, real-time modeling, human computer interaction, knowledge engineering and rule-based expert systems, and system simulation. He has held and participated in several funded research grants and contracts.